\pdfoutput=1

\documentclass[acmsmall]{acmart} 
\renewcommand\footnotetextcopyrightpermission[1]{} 
\AtBeginDocument{%
  }
    
\usepackage{algorithm}
\usepackage{algorithmic}

\usepackage{utfsym}

\usepackage{tabularx} 
\usepackage{array}

\usepackage{multirow} 
\usepackage{booktabs} 
\usepackage{makecell} 
\usepackage{array} 
\usepackage{multirow} 

\usepackage{ragged2e} 

\usepackage{caption}
\usepackage{array} 
\usepackage{cellspace} 
\usepackage{array} 
\usepackage{cellspace} 
\usepackage{colortbl} 
\definecolor{lightgray}{gray}{0.8} 

\usepackage[utf8]{inputenc}
\usepackage{listings}
\usepackage{xcolor}

\begin{document}

\title{Penetration Testing for System Security: Methods and Practical Approaches}

\author{Wei Zhang}
\authornotemark[1] 
\affiliation{%
  \institution{Hainan University}
  \city{Haikou}
  \country{China}}

\author{Ju Xing}
\authornote{These authors contributed equally to this work.}
\affiliation{%
  \institution{Hainan University}
  \city{Haikou}
  \country{China}}

\author{Xiaoqi Li}
\affiliation{%
  \institution{Hainan University}
  \city{Haikou}
  \country{China}}
\email{csxqli@ieee.org}

\begin{abstract}

Penetration testing refers to the process of simulating hacker attacks to evaluate the security of information systems . In this context, the paper introduces the fundamental concepts and overall procedures involved in penetration testing, while also analyzing the commonly used techniques and tools. Furthermore, it discusses the critical role and significance of penetration testing in the broader field of network security.  The research aims not only to clarify the theoretical foundations of penetration testing but also to explain and demonstrate the complete testing process, including how network system administrators may simulate attacks using various penetration testing methods. Methodologically, the paper outlines the five basic stages of a typical penetration test: intelligence gathering, vulnerability scanning, vulnerability exploitation, privilege escalation, and post-exploitation activities. In each phase, specific tools and techniques—such as Nmap and Metasploit—are examined in detail, along with practical guidance on their use.  To enhance the practical relevance of the study, the paper also presents a real-life case study, illustrating how a complete penetration test is conducted in a real-world environment. Through this case, readers can gain insights into the detailed procedures and applied techniques, thereby deepening their understanding of the practical value of penetration testing. In addition, the study analyzes common challenges and corresponding solutions encountered during testing, and provides guidance on how to effectively interpret and report test results. Finally, the paper summarizes the importance and necessity of penetration testing in securing information systems and maintaining network integrity, and it explores future trends and development directions for the field. Overall, the findings of this paper offer valuable references for both researchers and practitioners, contributing meaningfully to the improvement of penetration testing practices and the advancement of cybersecurity as a whole.

\end{abstract}

\keywords{System Security; Penetration Testing; Network Security; Information Security}
\maketitle

\pagestyle{plain} 

\section{Introduction}

With the continuous development of technology, computer communication and network information technology have expanded the use of digital services. As a historical indicator of this expansion, the 43rd Statistical Report on Internet Development in China reported 829 million Internet users and an Internet penetration rate of 59.6\% at the end of 2018, an increase of 3.8 percentage points over the end of 2017\cite{CNNIC43}. The growth of networked services increases the importance of protecting information and evaluating system security. Attackers may exploit software vulnerabilities and misconfigurations to gain unauthorized access, disclose or modify information, or disrupt services\cite{FIPS199,NIST800115}. Managing these risks requires organizations to define security objectives and select assessment techniques appropriate to their systems. Penetration testing provides an adversarial perspective on weaknesses that can be exploited under an agreed testing scope\cite{AboutPene,OverviewPentest}. This paper seeks to offer penetration testers a practical and intuitive approach to penetration testing. Specifically, it introduces an experimental testing platform and tool framework that can serve as a reference for practitioners. Through this research, our objective is to explain how penetration testing findings can inform the protection of enterprise information systems. Furthermore, we discuss methods, tools, and research directions in penetration testing, whose selection depends on the objectives and constraints of each assessment\cite{PenetrationTaxonomy}.

Threat intelligence can complement security assessment by providing indicators of compromise, attacker tactics and techniques, and incident analysis. Sharing this information can help organizations identify, assess, monitor, and respond to threats; it also requires explicit sharing objectives and rules for handling information\cite{NIST800150}.

With the continuous advancement of information technology, security assessment must consider the services exposed by increasingly complex networks. Reconnaissance helps identify reachable hosts, services, and potential attack surfaces. Cyber scanning research distinguishes scanning strategies and techniques, providing a basis for understanding the information gathered during an assessment\cite{CyberScanning}. At a larger scale, ZMap demonstrates how high-speed scanning can support Internet-wide security measurement\cite{ZMap}. These studies concern reconnaissance and measurement; they do not establish that scanning alone can prevent exploitation or eliminate unknown vulnerabilities.

Automation is another direction in penetration testing research. Earlier work surveys automated penetration testing approaches\cite{AutoPene}, while subsequent studies explore the use of large language models to assist security testing\cite{GettingPwn}. PentestGPT evaluates LLMs on penetration testing tasks and introduces a framework that addresses difficulties in maintaining the overall testing context\cite{Pentestgpt}. More recent work investigates agent-based testing and benchmarks for evaluating automated penetration testing\cite{AutoPen,TowardsAutomated}. These approaches motivate research on automation, but their reported results should not be treated as evidence for the performance of the manual network experiment in this paper.

Penetration testing research covers testing methods, tool capabilities, and the relationship between vulnerability discovery and practical exploitation\cite{AnOverofPene,AStudyonPene}. Automated scanning is useful, but it has coverage limitations. For example, an evaluation of black-box web vulnerability scanners found that application crawling is a major obstacle to discovering vulnerabilities\cite{WhyJohnnyPentest}. Software penetration testing should therefore complement other security assurance activities rather than provide a guarantee that a system is free of security defects\cite{SoftPene}. The following topics summarize research areas relevant to the testing workflow discussed in this paper.

\begin{itemize}
    \item Research on Penetration Testing Methods and Processes: Studies focus on using different methods and processes for penetration testing to provide better testing services and recommendations to enterprises.
    \item Research on Penetration Testing Tools: Research on the use of various tools to identify vulnerabilities and weaknesses in systems or applications. Many penetration testing tools are available on the market, such as Metasploit, Nmap, and Burp Suite.
    \item Research on Social Engineering: Social engineering is a technique used in penetration testing that exploits human vulnerabilities, such as trust and curiosity, to attack systems. Research on social engineering helps companies understand and protect themselves from such attacks.
    \item Vulnerability Discovery Research: Research on identifying vulnerabilities in systems or applications to improve the efficiency and precision of penetration testing. Vulnerability assessment and penetration testing provide complementary views of system weaknesses\cite{VulAssess}.
    \item Cloud Security Research: As enterprises increasingly adopt cloud services, the research on performing penetration testing in cloud environments becomes more important. This includes evaluating cloud security and using penetration testing tools and methods in the cloud.
\end{itemize}

Research on autonomous security analysis\cite{AutoSec} and reinforcement learning for network penetration testing\cite{ReinforceLearn} explores ways to automate assessment decisions. LLM-assisted reconnaissance has also been proposed\cite{MaximizingPene}. These are related approaches rather than components evaluated in the experiment reported here.

\section{Background}

\subsection{ Security of Network Information}

Information is a complex and abstract concept that can be expressed in various forms such as charts, data, symbols, numbers, programs, videos, etc. Different forms of information have different formats, encoding, syntax, and other structural types. Information refers to the content transmitted during the conversion process from one form or state to another. To ensure the security of information, it is essential to first understand its characteristics and then implement targeted security precautions.

In computer science, information is represented and processed through applications, storage media, and communication systems. Security assessment should focus on how these systems protect information against unauthorized access, disclosure, modification, and disruption. The confidentiality, integrity, and availability objectives defined in FIPS 199 provide a basis for specifying the security properties to be assessed\cite{FIPS199}.

Protecting information requires controls appropriate to the confidentiality, integrity, and availability objectives of the system. In particular, integrity concerns protection against improper modification or destruction of information\cite{FIPS199}. Security testing examines the implementation of relevant controls and uses the findings to guide mitigation\cite{NIST800115}.

Network information security protects network devices, applications, and information against unauthorized access and other threats. It combines technical controls with management processes throughout system development and operation. Testing activities may examine configuration, identify exposed services, scan for vulnerabilities, and assess whether discovered weaknesses can be exploited\cite{NIST800115}. The scope of information security also includes application, operating system, data, and physical security.

The goal of information security is to protect the confidentiality, integrity, and availability of information. Confidentiality concerns restrictions on access and disclosure; integrity concerns improper modification or destruction; and availability concerns timely and reliable access to information\cite{FIPS199}. Controls such as authentication, access control, and configuration management should be selected in response to the system's requirements. Their effectiveness must be evaluated within a defined scope rather than assumed from the presence of a security product\cite{NIST800115}.

Information security is a constantly developing and changing field. With the emergence and continuous improvement of new technologies, information security faces new challenges and threats. Therefore, maintaining information security requires continuous research and improvement to cope with ever-changing threats.

Security assessment methods must also reflect the application domain. In mobile application security, Li et al. survey Android vulnerabilities and approaches to discovering them\cite{OnDiscover}; Zhang et al. implement VScanner, combining exploit-based dynamic detection with feature matching for system-level vulnerabilities\cite{AnAndroidVul}. Chen et al. demonstrate system-level attacks against Android by exploiting asynchronous programming constructs\cite{SystemlevelAttack}. These studies illustrate why application-specific vulnerability analysis is relevant to security assessment, although the experiment in this paper focuses on a conventional network environment.

Blockchain applications provide a further example of domain-specific analysis\cite{SCLA}. SCALM uses large language models to identify bad practices in smart contracts\cite{Scalm}, while DeFiTail examines DeFi protocols through cross-contract execution analysis\cite{DeFiTail}. Opcode vectorization has also been studied for detecting malicious smart-contract code. At the transaction level, GasTrace analyzes transaction and gas-related features to detect sandwich-attack accounts\cite{GasTrace}; a cryptocurrency security survey organizes threats, attacks, and detection and defense approaches\cite{SoK}. These works concern smart-contract or transaction analysis and provide context for specialized assessment methods. They do not supply evidence for the network scanning results reported below.

\subsection{Penetration Testing }

Penetration testing is a security assessment technique in which authorized testers attempt to identify and exploit weaknesses within an agreed scope. NIST SP 800-115 describes penetration testing as an assessment activity involving attempts to circumvent the security features of a system\cite{NIST800115}. Test scope can include network services, operating systems, applications, and other assets, with the choice of methods determined by the objectives and available information\cite{OverviewPentest,PenetrationTaxonomy}.

Information gathering can include public information about the target organization, its applications, and its network infrastructure. For web applications, the OWASP Web Security Testing Guide describes information-gathering activities such as identifying the web server, application entry points, and application architecture\cite{OWASP42}. Such reconnaissance informs subsequent testing, but discovering a public resource does not itself demonstrate an exploitable vulnerability.

Penetration testing requires practical operation as well as planning and design. NIST SP 800-115 presents a four-stage example comprising planning, discovery, attack, and reporting, and explains that other groupings of the testing activities are also acceptable\cite{NIST800115}. The workflow used in this paper further distinguishes information gathering, vulnerability scanning, exploitation, and post-exploitation activities. Assessments must account for possible disruptions, define rules of engagement, and document findings and recommended mitigation\cite{NIST800115,ProfessionPene}. Application-specific testing has also been investigated in connected households, IoT devices, and the Robot Operating System\cite{PeneTestConnect,PENTOS,PeneTestROS}. These settings illustrate the need to tailor a testing workflow to its target.

\section{Standard Framework of Penetration Testing}

\subsection{Open Source Security Testing Methodology Manual}

The Open Source Security Testing Methodology Manual (OSSTMM) provides a methodology for testing operational security across five channels: human, physical, wireless, telecommunications, and data networks\cite{OSSTMM3}. It specifies the scope, targets, test vectors, and controls to be examined, and emphasizes recording what was tested and what remains outside the assessment. This approach supports the documentation of security testing across different types of organizational assets.

OSSTMM 3 organizes methodology execution into four phases: induction, interaction, inquest, and intervention\cite{OSSTMM3}. Its rules of engagement address test arrangements and responsibilities, while the Security Test Audit Report (STAR) records the scope and assessment findings. Applying these elements requires adaptation to the assets and constraints of the organization being assessed. The methodology can support the following activities:

\begin{itemize}
    \item  OSSTMM provides a structured approach for documenting and conducting security tests.
    \item  OSSTMM covers a range of security testing scenarios, including network security, application security, and physical security, meeting different security testing needs.
    \item  OSSTMM provides a structure for documenting test coverage, observed controls, and assessment findings. 
    \item  OSSTMM helps testers identify security vulnerabilities and risks, enabling them to address potential security issues during testing.
\end{itemize}

\subsubsection{Information System Security Assessment Framework} 

The Information Systems Security Assessment Framework (ISSAF) is a security assessment framework discussed alongside OSSTMM, NIST guidance, OWASP, and PTES in penetration testing literature\cite{PenetrationTaxonomy}. It addresses assessment activities across networks, systems, and applications, with attention to planning, technical assessment, and reporting. Its usefulness depends on selecting the parts that are appropriate to the test scope. Its intended functions include the following:

\begin{itemize}
    \item ISSAF is a standardized security assessment framework, It provides a consistent assessment method, enhancing comparability and repeatability.
    \item The developers of ISSAF have summarized a complete assessment framework from years of security evaluation practice, catering to diverse security evaluation needs.
    \item ISSAF provides an assessment structure; the reliability of an individual assessment depends on its implementation and coverage.
    \item ISSAF offers detailed guidance, reducing the difficulty and complexity of the assessment process.
\end{itemize}

\subsubsection{Techniques Guide of Information Security Testing and Evaluation }

NIST SP 800-115, \textit{Technical Guide to Information Security Testing and Assessment}, was published by the National Institute of Standards and Technology in September 2008 and superseded NIST SP 800-42\cite{NIST800115}. It presents key elements of technical security testing, the benefits and limitations of specific techniques, and recommendations for their use. Its purpose is to assist with planning assessments, analyzing findings, and developing mitigation strategies; it is not intended to provide a comprehensive assessment program.

The document covers a wide range of essential topics, including specific security tools, rules of engagement (RoE), and ethical guidelines that must be followed during testing. It outlines the necessary steps for security testing strategies, ensuring that testers have a clear roadmap for identifying vulnerabilities and assessing potential threats to a system. The framework also includes a detailed approach to role management, clarifying the responsibilities of different individuals involved in the penetration testing process.

\subsubsection{Open Web Application Security Project}

The OWASP Web Security Testing Guide (WSTG) provides guidance for testing web applications. Version 4.2 covers information gathering, configuration and deployment management, identity management, authentication, authorization, session management, input validation, error handling, cryptography, business logic, and client-side testing\cite{OWASP42}. It also discusses integrating security testing into development activities and reporting findings. Its testing scenarios can be selected and adapted according to the application and test objectives.

\subsection{Standard of Penetration Testing Execution}

The Penetration Testing Execution Standard (PTES) organizes a penetration test into seven main sections: pre-engagement interactions, intelligence gathering, threat modeling, vulnerability analysis, exploitation, post-exploitation, and reporting\cite{PTES}. Its accompanying technical guidelines provide further details. The seven sections support the following activities:

\begin{enumerate}
    \item Pre-engagement Interactions: Agree on objectives, scope, authorization, communication arrangements, and rules of engagement.
    \item Intelligence Gathering: Collect information about the organization and the assets included in the test scope.
    \item Threat Modeling: Identify relevant threats and prioritize plausible attack scenarios.
    \item Vulnerability Analysis: Analyze discovered services, applications, and configurations for potential weaknesses.
    \item Exploitation: Attempt to exploit selected weaknesses within the agreed scope to determine their practical impact.
    \item Post-exploitation: Assess the value of compromised assets and the consequences of the access obtained, subject to the rules of engagement.
    \item Reporting: Document the assessment, evidence, impact, and recommended remediation.
\end{enumerate}

\section{Platforms and Tools of Penetration Testing }

\subsection{Kali Linux}

Kali Linux is a Debian-based Linux distribution primarily intended for security professionals and penetration testers\cite{KaliDocs}. Developed and maintained by Offensive Security Ltd, it is a free and open-source project that comes preloaded with a comprehensive array of security tools. These tools span multiple domains, including network penetration testing, vulnerability analysis, reverse engineering, wireless attacks, password cracking, and digital forensics. Its accessibility and flexibility make it a go-to platform for security practitioners worldwide, as it can be freely downloaded and used without any restrictions.

A key strength of Kali Linux lies in its extensive collection of built-in security tools. Notable examples include Metasploit, Nmap, Wireshark, John the Ripper, Aircrack-ng, Hydra, and Burp Suite. These tools allow security professionals to efficiently identify and exploit vulnerabilities, facilitating deeper penetration testing and more comprehensive assessments. Designed specifically for advanced penetration testing and security auditing, Kali Linux provides a complete toolkit that serves the needs of security researchers, penetration testers, digital forensic analysts, and reverse engineers. This broad and powerful toolset is one of the main reasons Kali Linux was selected as the foundation for this research. Its latest release already includes the majority of tools discussed in this paper. Built in accordance with Debian Linux development standards, Kali Linux ensures high stability and compatibility, and it minimizes common bugs that users might encounter during installation or tool usage.

In addition to its technical capabilities, Kali Linux is highly customizable, allowing users to tailor the environment by adding or removing tools based on specific testing requirements. It also supports multiple desktop environments such as GNOME, KDE, and Xfce, offering flexibility in user interface preferences. While it is an indispensable platform for security professionals and penetration testers, Kali Linux is equally well-suited for students and researchers engaged in cybersecurity education, academic studies, and hands-on lab experiments. In summary, Kali Linux is a robust, security-focused operating system that offers a wide range of tools and strong adaptability, enabling users to conduct penetration testing tasks more efficiently. The key features of Kali Linux can be summarized as follows:

\begin{itemize}
    \item More than 600 Penetration Testing Tools and totally free to user: All tools from BackTrack were reviewed; redundant or non-functional ones were remoOpen Sourceved.
    \item Open Source: Kali follows an open development model. Its development tree is publicly accessible, and all source code can be modified and rebuilt based on specific needs.
    \item Secure Development Environment: Only trusted developers under strict security protocols are authorized to interact with core libraries and submit updates.
    \item ARMEL and ARMHF Support: With the rising popularity and affordability of ARM-based systems like Raspberry Pi and BeagleBone Black, Kali Linux offers strong ARM support. ARM builds are released alongside standard versions and include a dedicated ARM repository, ensuring excellent performance on both ARMEL and ARMHF platforms.
\end{itemize}

\subsection{Metasploit}

Metasploit Framework is an open-source penetration testing framework used to investigate and validate vulnerabilities. Its modules support exploitation and related testing activities\cite{MetasploitDocs}.

The Metasploit framework consists of three core components: Metasploit Framework, Metasploit Pro, and Metasploit Community Edition. The Metasploit Framework is the foundational and most widely used component, offering a comprehensive suite of tools and modules for tasks such as vulnerability scanning, exploitation, payload generation, shellcode development, and post-exploitation.

Core Components of Metasploit Framework are concluded as follows: 
\begin{itemize}
    \item Module Library: A rich repository that includes modules for vulnerability scanning, exploitation, payload generation, shellcode writing, and post-exploitation.
    \item Interfaces: Supports various interfaces including command-line interface (CLI), graphical user interface (GUI) through Armitage, and web-based interfaces.
    \item Database Integration: Compatible with databases such as PostgreSQL, MySQL, and SQLite for storing and managing data.
    \item Payloads: Offers a variety of payloads for executing actions on target systems, such as gaining shell access, uploading/downloading files, and executing commands.
    \item Exploits: Provides numerous exploit modules for leveraging system vulnerabilities to gain shell access.
    \item Auxiliary Modules: Includes modules for auxiliary tasks such as port scanning, service identification, and vulnerability verification.
\end{itemize}

The Metasploit Framework is highly flexible, allowing users to select and configure different modules and exploits tailored to specific targets. Additionally, users can develop custom modules and exploits to suit particular testing scenarios, enhancing the tool’s adaptability to diverse environments and security challenges.

Before launching an attack using Metasploit, penetration testers must first gather detailed information about the target. Metasploit contains a vast database of vulnerabilities related to applications, protocols, services, and operating systems. Once a vulnerability is identified, testers select an appropriate exploit module. If successful, the exploit allows the tester to execute a payload module on the target machine. These payloads can return a shell or establish a connection between the target and the tester’s machine, thereby enabling remote command execution and further system interaction.

\begin{figure}[h]
    \centering
    \includegraphics[width=0.8\textwidth]{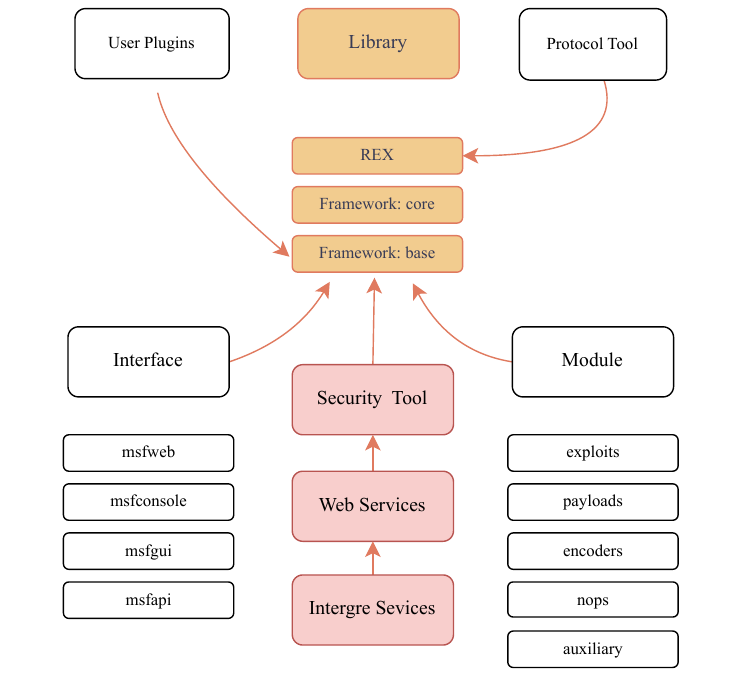}
    \caption{Metasploit Architecture }
    \label{fig:Metasploit Architecture}
\end{figure}

Before executing an exploit, the payload must be pre-configured properly. To avoid detection by intrusion prevention and detection systems, payloads can be encoded. Metasploit supports automation of the entire exploitation process through appropriate scripting and configuration, streamlining workflows and improving testing efficiency. In summary, Metasploit Framework is a powerful, user-friendly, and extensible penetration testing tool—an essential platform for vulnerability research, exploitation, and security testing. Fig. \ref{fig:Metasploit Architecture} illustrates the Metasploit architecture.

Metasploit includes various subdirectories for interacting with the core framework, which are categorized into interfaces, modules, libraries, plugins, and tools. REX is an extended Ruby library that integrates classes and methods within the Metasploit base files. The interfaces provide five main options: msfweb, msfcli, msfconsole, msfgui, and msfapi, among which msfconsole is the most powerful, granting users full access to all Metasploit features. Additionally, the plugin interface allows for extending the framework’s architecture and capabilities. The true power of Metasploit lies heavily in its modularity, as each module serves a distinct and specific purpose within the penetration testing process.

\subsection{Nmap}

Nmap is a network exploration and security auditing tool. Its official reference guide documents host discovery, port scanning, service detection, operating system detection, and related options\cite{NmapGuide}. These capabilities make it useful for identifying the network services included in the scope of a penetration test.

Nmap’s core functions include scanning hosts, services, operating systems, packet filters, and firewalls. It features advanced capabilities that allow it to bypass firewalls and intrusion detection systems. In addition, Nmap offers numerous hidden functions that can be integrated into scripts and other programs. Alongside these, it also comes with a suite of auxiliary tools, for example, Ncat for debugging, Ndiff for comparing scan results, and Nping for packet generation and response analysis, among others.

Regarding output, Nmap presents scan results in a list format based on the selected options. For example, the port list displays the port number, protocol, service name, and status, with statuses that include open, filtered, closed, unfiltered, and undefined. Specifically, "open" indicates that Nmap has detected a service on the target host, while "filtered" suggests that a firewall or other network barrier is blocking the port, preventing Nmap from receiving a response. Moreover, the Nmap Scripting Engine (NSE) is a powerful framework designed to perform a wide range of network-related tasks, including network discovery, advanced version detection, vulnerability scanning, backdoor detection, and even vulnerability exploitation. Users can leverage NSE to automate various operations, which are executed in parallel while optimizing performance and speed. In addition to the built-in scripts, users can develop custom scripts tailored to specific requirements. These custom scripts are particularly valuable in penetration testing, enabling more targeted and efficient assessments.

\subsection{OpenVAS}

OpenVAS Scanner is the scan engine used in the Greenbone vulnerability management ecosystem. It executes a continuously updated feed of vulnerability tests and inspects target hosts for vulnerabilities and common misconfigurations\cite{GreenboneDocs}. The available tests and scan configuration determine the coverage of an assessment. This study uses OpenVAS for the vulnerability scanning stage of the network experiment.

A key feature of OpenVAS is its ability to use plugins to match and identify vulnerabilities effectively.

\begin{figure}[h]
    \centering
    \includegraphics[width=0.8\textwidth]{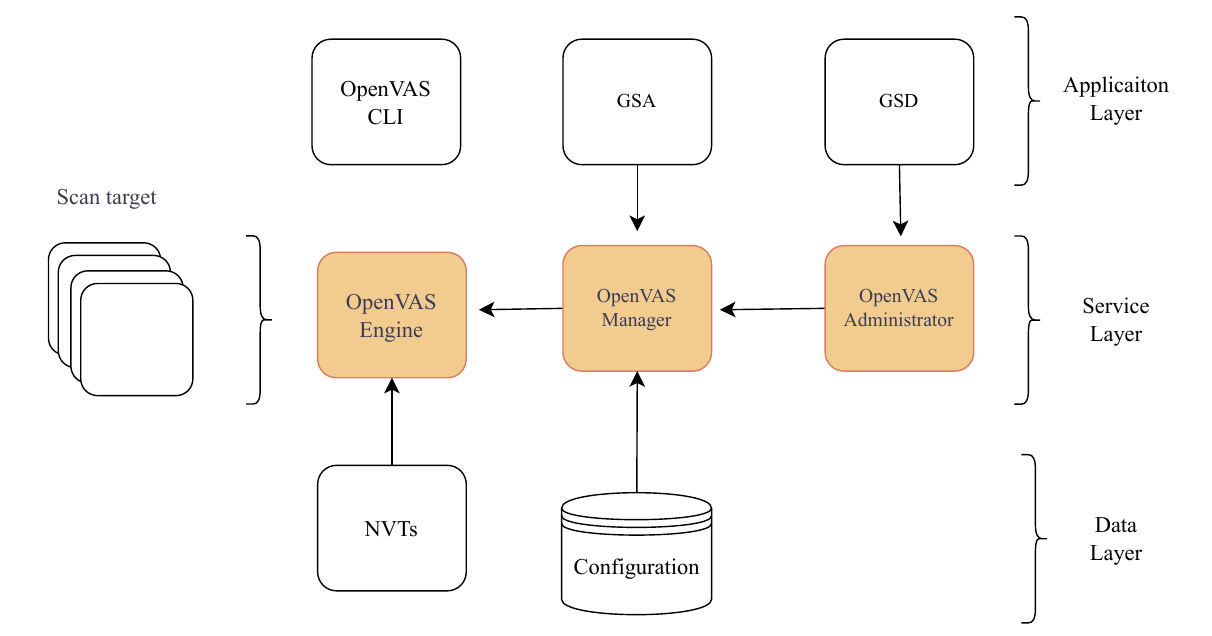}
    \caption{The Three Layers of OpeVAS}
    \label{fig:The Three Layers of OpeVAS}
\end{figure}

The architecture of OpenVAS can be divided into three main layers: the client layer, the service layer, and the data layer. The essential components and modules within these layers are illustrated in Fig. \ref{fig:The Three Layers of OpeVAS}

VirtualBox is a hypervisor that enables the operation of a virtual operating system within an existing host OS. It has continuously evolved, offering new features and improvements. Its user interface is based on QtGUI, and it includes a command-line management tool to operate VBoxSDL. To integrate the host system’s capabilities into the virtual environment, VirtualBox provides an additional software package that includes shared folders and clipboard support, video acceleration, and a seamless window experience.

VirtualBox supports a wide range of operating systems, including x86 and AMD64/Intel64 architectures. It is currently the only open-source virtual machine software licensed under the GNU General Public License (GPL) version 2.0. It runs smoothly on multiple host systems such as Linux, Windows, macOS, and Solaris, and supports a variety of guest operating systems including Windows, Linux, Solaris, and OpenBSD. With continued development, VirtualBox is expected to support more features and a broader range of operating systems and platforms.

Funded by Oracle, VirtualBox is a community-driven open-source project to which anyone can contribute. Oracle ensures that the product maintains professional quality at all times.

Compared to VMware, VirtualBox is arguably a better choice for general users. For cybersecurity researchers, setting up a targeted, isolated network is an effective strategy. In this project, VirtualBox is used to build a small-scale test network entirely independent of real-world applications, ensuring that penetration tests can be conducted without affecting any actual networks. VirtualBox offers four distinct networking modes to accommodate different user networking requirements.

\section{Experiments}

\subsection{ Experimental Platform Design }

\begin{figure}[h]
    \centering
    \includegraphics[width=0.5\textwidth]{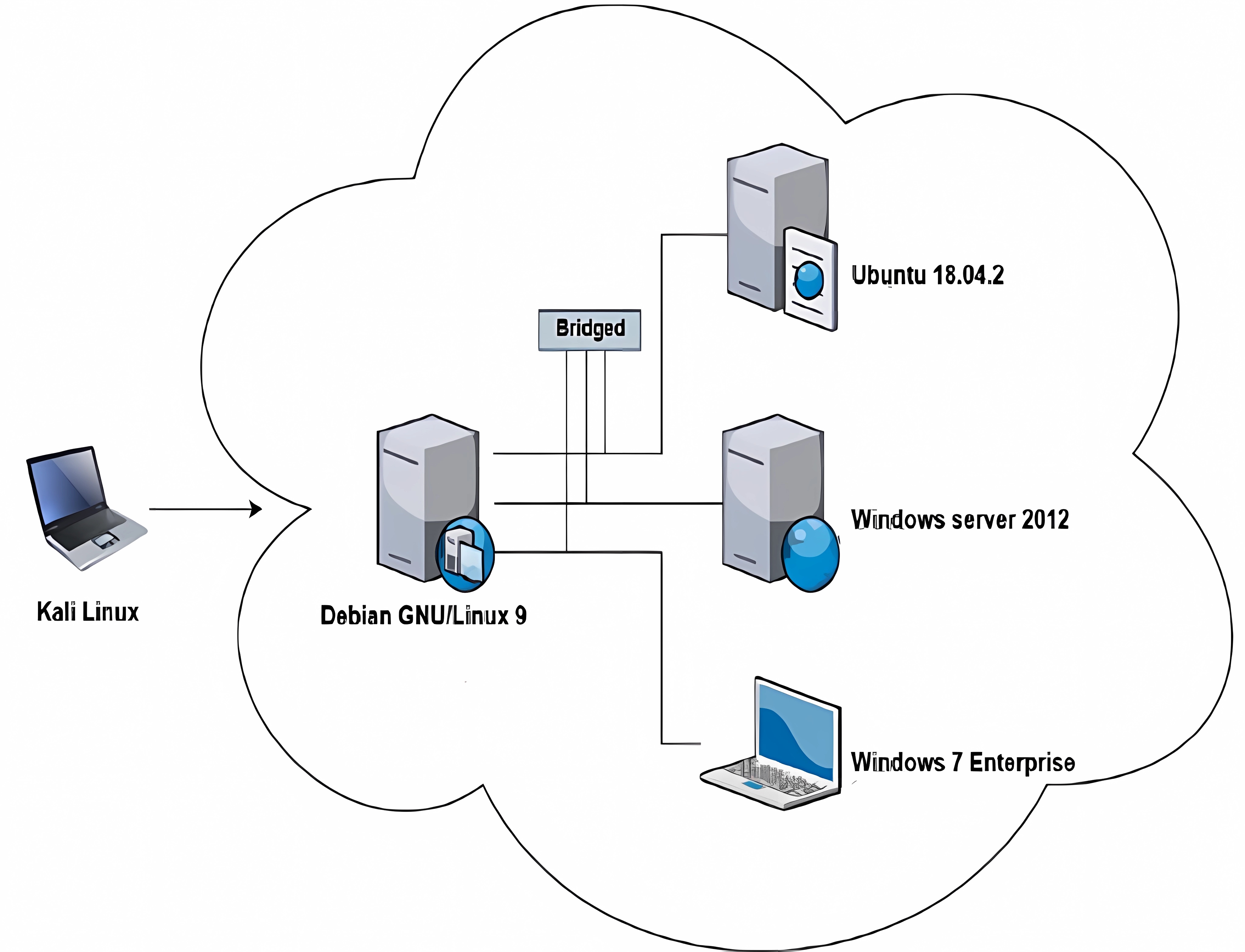}
    \caption{Enviroment of Platform}
    \label{fig:Enviroment of Platform}
\end{figure}

Due to legal and time constraints, it is essential to separate the experimental environment from any real production systems. To achieve this, two standalone computers were used for testing. The system was designed to be entirely independent of any external network devices, allowing full control by the user. One machine served as the attacker running the penetration tests, with Kali Linux installed as the operating system. The other machine used VirtualBox to create a virtual LAN, which served as the target environment for testing. This virtual LAN simulates an organized network comprising multiple servers and clients, reflecting the structure of a real-world network to a certain scale, and the platform enviroment is illustrated in Fig. \ref{fig:Enviroment of Platform}

The machine used for penetration testing runs exclusively on Kali Linux. Selecting an appropriate ISO image during installation is crucial. Since malicious software such as viruses or trojans may be embedded in downloads from the internet, verifying the integrity of the image is an important step to ensure it has not been tampered with or corrupted. For instance, in Windows, the certutil -hashfile command can be used, while in Linux, tools like md5sum or sha256sum serve the same purpose. This verification helps avoid installation failures caused by packet loss during download and minimizes potential security risks. Once the image is confirmed to be intact, the operating system can be installed with confidence.

The target network simulates the internal network of a small organization. Within VirtualBox, a DHCP server, FTP server, web server, and client PCs were created, each installed with different operating systems: Debian, Ubuntu, Windows Server 2012 R2, and Windows 7 Enterprise. These systems were organized into a virtual local network. This setup helps penetration testers understand the security differences across various platforms.

A well-structured hardware configuration significantly enhances the efficiency of penetration testing. However, due to the differing scopes of experimental and production environments—as well as limitations imposed by legal and financial factors—moderate-performance hardware was selected. Kali Linux offers broad hardware compatibility through its well-developed drivers, eliminating the need for high-end hardware on the attacking host. In contrast, the target system is required to simulate multiple virtual devices, necessitating the use of a multi-core processor and large-capacity memory to ensure smooth execution of penetration tests. 

\subsection{Design of the Experimental Process}

In practical applications, penetration testing plays a critical role. However, it also carries considerable risk, potentially leading to system or network failures and even economic loss. To better understand penetration testing techniques and methodologies, this study proposes an experimental process tailored for penetration testing.

\begin{figure}[h]
    \centering
    \includegraphics[width=0.8\textwidth]{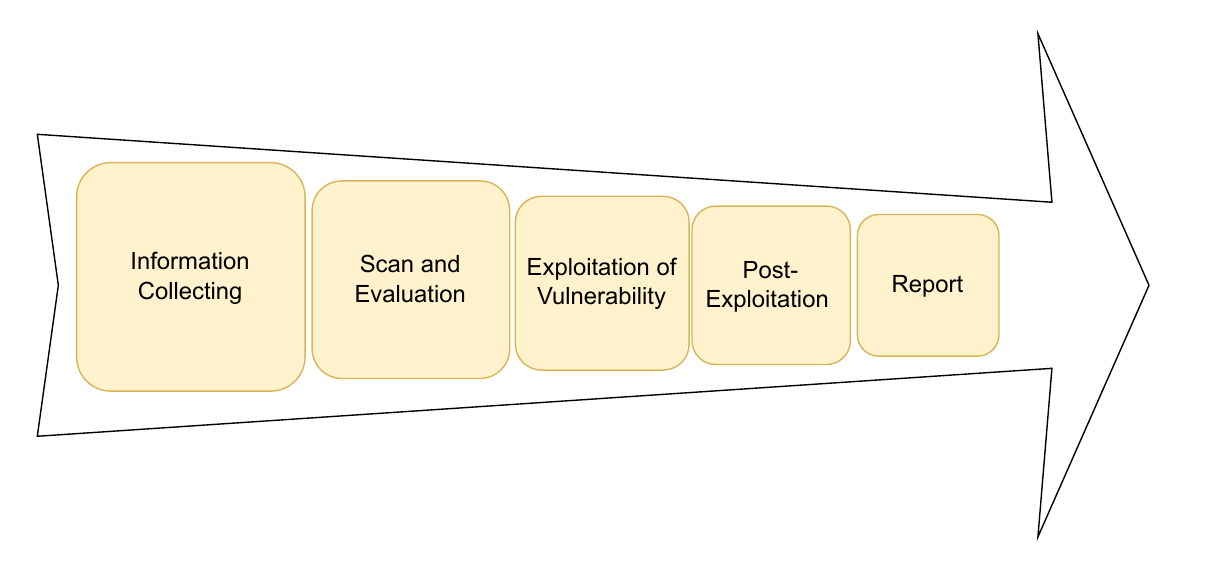}
    \caption{Steps of Penetration Test}
    \label{fig:Steps of Penetration Test}
\end{figure}

Based on this, a scenario-oriented penetration testing method is introduced, with the detailed steps illustrated in Fig. \ref{fig:Steps of Penetration Test}

In the information collection phase, Nmap was used for network reconnaissance, port scanning, OS detection, and service enumeration. In practice, information gathering can consume over 60\% of the total testing time, highlighting its critical role in the overall process. The quality of the collected data significantly influences the effectiveness of subsequent stages. Specifically, data such as network coverage, host IP addresses, operating system versions, and service ports serve as key indicators in the following vulnerability assessment phase.

Moving on to the vulnerability scanning and evaluation phase, the collected data is carefully reviewed and refined. This process can be performed manually or with the assistance of automated tools. However, manual efforts tend to be time-consuming and inefficient, especially in large-scale networks consisting of hundreds of subsystems. In this study, the open-source tool OpenVAS was employed to detect vulnerabilities by leveraging its constantly updated vulnerability database, which helps identify issues caused by misconfigurations or other factors.

The exploitation phase follows, involving the verification and exploitation of vulnerabilities identified by OpenVAS. While often regarded as the most exciting part of penetration testing, it also carries the highest implementation risk. Many security vulnerabilities in real-world scenarios are difficult to exploit and may adversely affect system stability. Nevertheless, within a controlled testing environment, tools such as the Metasploit Framework can be safely used without concerns about damaging the target system.

Finally, post-exploitation refers to actions taken after successfully breaching a target network system, including deeper exploration and the installation of rootkits or backdoors to ensure persistent access. Kali Linux, combined with the Metasploit Framework, provides a variety of tools to support this process. Generally, penetration tests do not directly improve an organization’s security posture but instead provide recommendations through a detailed report. It is then the responsibility of the target organization to develop and implement specific security measures, balancing effectiveness and practicality based on the report’s findings. After completing all test steps, a comprehensive written report must be prepared and submitted, encompassing test results, recommended improvements, a list of tools used, and a summary of all procedures executed. This penetration testing report encapsulates the entire testing process and serves as a crucial guide for future security enhancements.

\subsection{Penetration Testing}

The information gathering phase is critical for analyzing and identifying weaknesses within an organizational network\cite{CyberScanning}. This phase is typically divided into three main categories: passive reconnaissance, active reconnaissance, and network/system scanning. Based on these approaches, tasks such as host discovery, port scanning, service version detection, and OS fingerprinting are conducted and analyzed\cite{NmapGuide}  . Notably, host discovery is rarely performed in isolation; it is generally used as a preparatory step for subsequent tasks such as port scanning, version detection, and OS identification. Only in specific scenarios—such as counting active switches in a large-scale network—does host discovery become a standalone operation. In this experiment, Nmap was used to scan a segment of a test network. The scan revealed the number of active hosts in the target subnet, along with details such as their IP addresses, MAC addresses, and associated network interface card (NIC) information . In practical network environments, ICMP traffic is often filtered by firewalls. To overcome the limitations of ICMP-based scanning, a combination of TCP and UDP scanning is typically recommended. However, because this combined approach is time-consuming, time constraints must be considered during penetration testing to avoid unnecessary delays and to ensure the efficiency of the overall testing process.

To address this, Nmap offers a wide range of command-line options to support various scanning techniques. In this experiment, the TCP ACK scan mode was used to simulate a gray-box testing scenario. Four VirtualBox hosts identified during the host discovery phase were selected as the targets. By performing TCP ACK scans on their IP addresses, we were able to infer their firewall configurations. The scan results indicated that one of the hosts (192.168.0.100) was in a "filtered" state, suggesting the presence of a firewall, whereas the remaining hosts were "unfiltered." TCP ACK scans help characterize packet filtering but do not establish whether ports are open or closed\cite{NmapGuide}. Therefore, additional scans using TCP SYN and UDP are necessary to obtain a more comprehensive understanding. Since UDP scanning is generally slower and more resource-intensive than TCP scanning, TCP SYN scanning was selected for this experiment.

In the subsequent vulnerability detection and assessment phase, OpenVAS was used to evaluate the target network. OpenVAS executes vulnerability tests against the configured targets\cite{GreenboneDocs}. OpenVAS is a reliable and effective vulnerability scanner, though its installation and configuration can be somewhat complex, as discussed earlier. This section provides a deeper exploration of OpenVAS and an analysis of the results produced during the experiment. To create a scan task with OpenVAS, configuration must be based on the information gathered in the previous phase. This involves specifying key parameters such as the IP address of the scan target, scan intensity, and task details. The number of hosts in the target network, the network's condition, the breadth of the vulnerability database, and the intensity of the scan all significantly influence both scanning efficiency and duration. In experimental scenarios, it is advisable to reduce the scan intensity and scope appropriately based on test conditions, thereby minimizing execution time. OpenVAS scan results provide valuable insights into the vulnerabilities and overall security posture of the target network. It generates detailed scan reports containing information such as IP addresses, user identities, scan time, scope, and types of detected vulnerabilities. These reports can be categorized according to various criteria, including operating system, IP address, or scan task. Penetration testers can also review detailed attack-related information based on the report format selected. During the scanning of the test network, two high-risk and two low-risk vulnerabilities were identified. These could potentially pose serious security threats. To enhance the accuracy of its detection, OpenVAS first determines the target system’s operating system and IP address before initiating a scan. The consistency of OpenVAS results with earlier reconnaissance findings further validates the effectiveness of the information gathering phase. Based on the analysis of scan outcomes, appropriate security measures can then be developed to improve system protection. In a test environment, it is possible to exploit high-risk zones for further analysis. However, in real-world penetration testing, exploiting critical vulnerabilities may interfere with normal system operations. For this reason, such actions should only be taken after thorough evaluation and approval by the penetration testing team. As part of the test, one of the identified vulnerabilities covered by the MS17-010 bulletin was examined in detail for its potential impact. Microsoft describes the affected SMBv1 vulnerabilities and associated security update in the original bulletin\cite{MS17010}.

\section{Conclusion}

Penetration testing refers to the practice of assessing the security of network systems by simulating real-world attacks \cite{PeneTestAHand}\cite{PeneTestConcept}. This paper presents an overview of the methods and practices of penetration testing and highlights its significance in the field of cybersecurity. It begins by introducing the fundamental concepts and classifications of penetration testing, with each type tailored to specific application scenarios and corresponding methodologies. To support practical implementation, the paper also discusses widely used tools and techniques—such as Nmap, Metasploit, and Burp Suite—that form the foundation of many penetration testing workflows\cite{PeneTestNetwork}\cite{EthicalHack}.  Building upon this theoretical basis, the paper then outlines the typical procedures involved in a penetration test, including information gathering, vulnerability scanning, exploitation, privilege escalation, data extraction, and post-test cleanup\cite{ProfessionPene}.  Each of these stages requires the use of different tools and technical approaches. The importance of penetration testing is subsequently analyzed in the broader context of network security, as it enables organizations to identify and remediate vulnerabilities, enhance system defenses, and reduce the risk of cyberattacks\cite{PeneTestConnect}.  Moreover, it assists in evaluating an organization’s current security posture, thereby facilitating the development of more effective security policies and strategic planning.

Expanding upon these foundational concepts, the study constructs a controlled test environment isolated from real-world production systems, allowing for experimentation with reduced exposure to operational networks. From the perspective of a penetration tester, the research utilizes multiple tools available on the Kali Linux platform to conduct in-depth testing on a simulated network. This includes comprehensive analysis of systems, network infrastructure, virtualization software, and specialized penetration testing tools. Based on the experimental results, the study proposes a set of practical implementation strategies aimed at addressing specific vulnerabilities and structural weaknesses within the tested network system. In addition, corresponding defensive measures are recommended to improve overall system security and resilience. Throughout the testing process, several issues were identified—ranging from software vulnerabilities to broader security configuration risks. By appropriately applying established penetration testing methodologies and leveraging suitable tools, these problems were successfully diagnosed and resolved. The proposed solutions serve to assist network administrators in better protecting their environments and proactively mitigating potential threats.

The results of this research demonstrate that, when systematically and effectively applied, penetration testing is a powerful means of identifying security flaws and implementing appropriate countermeasures. The findings contribute to both the theoretical understanding and practical advancement of cybersecurity, thereby reinforcing the reliability of networked systems. As cyber threats continue to evolve, penetration testing tools and techniques must also adapt to address emerging vulnerabilities and increasingly complex attack vectors.  In conclusion, penetration testing has become an indispensable component of modern network security. It plays a vital role in enhancing system protection and safeguarding sensitive information. Looking ahead, future work should prioritize the refinement of penetration testing methodologies to keep pace with the dynamic cybersecurity landscape. However, while penetration testing can inform the design of robust security architectures, it is also important to acknowledge that overly stringent security measures may interfere with the normal operation of systems. Therefore, a scientific and balanced approach is required—one that ensures both effective security assurance and the maintenance of operational efficiency.

\bibliographystyle{ACM-Reference-Format}
\bibliography{main}

\end{document}